\documentclass[pageno]{jpaper}

\usepackage{graphicx}
\usepackage{adjustbox}
\usepackage[normalem]{ulem}
\usepackage{svg}
\usepackage{array,multirow,graphicx}
\usepackage{pbox}
\usepackage{tikz}
\def\checkmark{\tikz\fill[scale=0.4](0,.35) -- (.25,0) -- (1,.7) -- (.25,.15) -- cycle;} 
\usepackage{xcolor}
\usepackage{soul}
\usepackage{tabularx}
\usepackage{amsmath}
\usepackage{hyperref}
\usepackage[numbers]{natbib}
\usepackage{url}
\begin{document}

\title{TAPA-CS: Enabling Scalable Accelerator Design on Distributed HBM-FPGAs}
\date{}
\author{\normalsize{Neha Prakriya, Yuze Chi, Suhail Basalama, Linghao Song, Jason Cong}\\ \normalsize{University of California, Los Angeles}\\ \normalsize{nehaprakriya, chiyuze, basalama, linghaosong, cong@cs.ucla.edu}}
\maketitle

\thispagestyle{empty}
\begin{abstract}

Despite the increasing adoption of Field-Programmable Gate Arrays (FPGAs) in compute clouds, there remains a significant gap in programming tools and abstractions which can leverage network-connected, cloud-scale, multi-die FPGAs to generate accelerators with high frequency and throughput. To this end, we propose TAPA-CS, a task-parallel dataflow programming framework which automatically partitions and compiles a large design across a cluster of FPGAs with no additional user effort while achieving high frequency and throughput. TAPA-CS has three main contributions. First, it is an open-source framework which allows users to leverage virtually "unlimited" accelerator fabric, high-bandwidth memory (HBM), and on-chip memory, by abstracting away the underlying hardware. This reduces the user's programming burden to a logical one, enabling software developers and researchers with limited FPGA domain knowledge to deploy larger designs than possible earlier. Second, given as input a large design, TAPA-CS automatically partitions the design to map to multiple FPGAs, while ensuring congestion control, resource balancing, and overlapping of communication and computation. Third, TAPA-CS couples coarse-grained floorplanning with automated interconnect pipelining at the inter- and intra-FPGA levels to ensure high frequency. We have tested TAPA-CS on our multi-FPGA testbed where the FPGAs communicate through a high-speed 100Gbps Ethernet infrastructure. We have evaluated the performance and scalability of our tool on designs, including systolic-array based convolutional neural networks (CNNs), graph processing workloads such as page rank, stencil applications like the Dilate kernel, and K-nearest neighbors (KNN). TAPA-CS has the potential to accelerate development of increasingly complex and large designs on the low power and reconfigurable FPGAs. On average, the 2-FPGA, 3-FPGA, and 4-FPGA designs are 2.1$\times$, 3.2$\times$, and 4.4$\times$ faster than the single FPGA baselines generated through Vitis HLS. The superlinear speed-up is due to the increased memory bandwidth and parallelization opportunities enabled by multiple FPGAs. For each benchmark, we also analyze the computation to communication trade-offs that enable efficient multi-FPGA design. Through intelligent floorplanning and interconnect pipelining, TAPA-CS achieves a frequency improvement between 11\%-116\% compared with Vitis HLS. Lastly, we discuss the challenges in scaling to multiple server nodes containing 8 FPGAs, and directions for future research in this area. 
\end{abstract}
\section{Introduction} \label{introduction}
In the big data era, there has been an exponential rise in the demand for scalable, cheap, and high performance acceleration. FPGAs have emerged as a promising solution to counter the breakdown of Dennard's scaling \cite{1050511, 10.1145/1941487.1941507} due to their reconfigurability and low power consumption. One of the greatest successful demonstrations is Microsoft's Catapult project which sped up the Bing Search engine using Stratix V FPGAs, achieving a 95\% increase in throughput with a minimal power consumption increase of only 10\% \cite{7106407}. Microsoft also displayed the use of FPGAs for accelerating DNN inference and data compression in their servers \cite{ovtcharov2015accelerating, fowers2015a, caulfield2016a, fowers2018a, chung2018serving}. Today, other major players such as Amazon \cite{aws, aqua}, Alibaba \cite{alibaba}, Baidu \cite{baidu}, and Huawei \cite{huawei} also use FPGAs to accelerate their workloads, and offer them as a service in their cloud. Most of these are achieved through manual RTL coding.

\begin{figure}
  \centering
  \includegraphics[height=5cm, keepaspectratio]{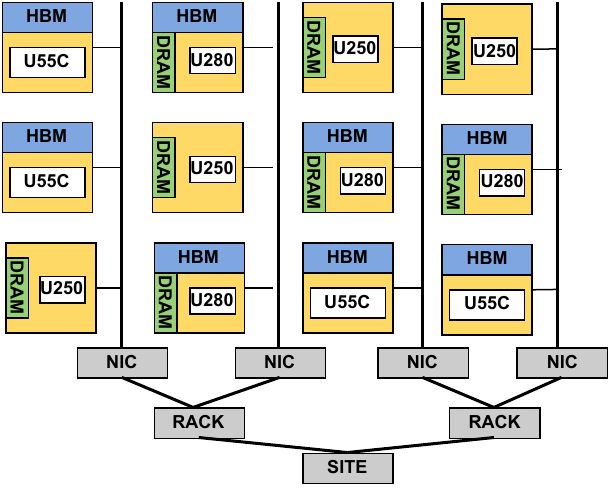}
  \caption{Network-Connected FPGAs}
  \label{fig:1}
\end{figure}
\begin{table*}[h]
\adjustbox{max width=\textwidth}{
\begin{tabular}{|c|c|c|c|c|c|c|c|c|c|}
\hline
Method & HLS & Ethernet & Floorplanning & Interconnect Pipelining & Topology-Aware & Automatic Partitioning & Hardware Execution & Generalizable & Fmax (MHz) \\
\hline
FPGA'12\cite{fleming} & $\times$ & $\times$ & $\times$ & $\times$ & $\times$ & $\times$ & $\times$ & $\checkmark$ & 85 \\
\hline
Simulation-based \cite{8416816, 9319909, 10.1145/2694344.2694362} & $\times$ & $\times$ & $\times$ & $\times$ & $\times$ & $\checkmark$ & $\times$ & $\checkmark$ & - \\
\hline
Virtualization-based \cite{10.1145/3373376.3378491, 9499729, 10.1145/3445814.3446699, 10.1145/3575693.3575753} & $\checkmark$ & $\checkmark$ & $\times$ & $\times$ & $\times$ & $\checkmark$ & $\checkmark$ & $\checkmark$ & 100-300\\
\hline
CNN/DNN \cite{alonso, cong, Baskin_2018, 8425399, 8715174, 9114652, weiwen} & $\checkmark$ & $\checkmark$ & $\times$ & $\times$ & $\times$ & $\checkmark$ & $\checkmark$ & $\times$ & 240 \\
\hline
\textbf{TAPA-CS (Ours)} & $\checkmark$ & $\checkmark$ & $\checkmark$ & $\checkmark$ & $\checkmark$ & $\checkmark$ & $\checkmark$ & $\checkmark$ & 300\\
\hline
\end{tabular}}

\caption{Comparison of TAPA-CS and existing methods providing scale-out acceleration across multiple FPGAs.}
\label{comparisons}
\end{table*}
High-level synthesis (HLS) tools like Vitis HLS \cite{vitis_hls} and Intel HLS \cite{intel_hls} raise the abstraction level for programming individual FPGAs in the cloud from RTL to C++/OpenCL. This allows the programmer to have little to no knowledge of the underlying hardware and cycle accuracy. While these tools deliver great results, they are limited to programming a single FPGA. At the same time, with the gradual end of Moore's law, accelerator designs are becoming larger than ever before, and require more programmable logic and memory than that available on a single FPGA device \cite{9499852}. Our goal is to support the network-connected devices shown in Figure \ref{fig:1}. Utilizing such multi-FPGA setups requires careful consideration for high-speed communication support and efficient workload distribution to ammortize the cost of inter-FPGA communication. 

The design of modern FPGAs also adds a new layer of complexity. Modern FPGA architectures have varying interconnection support (PCIe and Ethernet-based QSFP28 ports), different types of memory bandwidth and capacity (on-chip BRAM, off-chip DRAM, and HBM), programmable logic units organized into multiple dies (i.e., chiplets), and degrees of data transfer cost (on-die, cross-die, cross-chip). Consider the example of the Xilinx/AMD Alveo U55C cards. This card supports 2 Ethernet-compatible QSFP28 ports for networking, offering 100GBps bandwidth per port. It also features an HBM with 16GB capacity exposing a bandwidth of 460GBps. The on-chip memory provides a high bandwidth of 35TBps but has a small capacity of 43MB \cite{u55c}. Furthermore, these FPGAs are divided into multiple dies joined by silicon interposers, with a high inter-die crossing delay. We introduce details of modern FPGA architecture in Section \ref{background}.

An expert designer will consider all these factors when designing and partitioning their kernel code across chips. However, manual workload partitioning is inefficient as the design gets larger, and, in-turn, raises the barrier for using FPGAs. Therefore, despite the advances in CAD tools which allow the user to program a single \textit{FPGA in the cloud}, there is a significant lack of programming tools which can target \textit{multiple FPGAs} potentially at the cloud scale. Such a framework should take as input large workloads from the user and automatically partition it across multiple devices efficiently. We identify the following three main challenges to address when developing such frameworks for cloud-scale FPGAs:

\begin{enumerate}
  \item Need a lightweight inter-FPGA communication infrastructure which enables reliable and high speed data transfers.
  \item Need to partition and map application code efficiently keeping in mind factors like compute-load balancing, network topology, the varying cost of on- and off-chip communication, and keeping resource utilization in each die under a specified threshold. 
  \item Need to ensure high design frequency by hiding inter-FPGA communication latency, and sufficiently pipelining the interconnect.
\end{enumerate}

Several prior works have attempted to leverage the networking capabilities available in modern FPGAs \cite{4100995, 8541106, 10.1145/3295500.3356201, inproceedings, 9114837, 9651265, alveolink}. These prior works differ in the achievable data transfer throughput (10-90GBps), orchestration of data transfers (host/FPGA), and the resource overheads. We compare these methods in detail in Section \ref{prior_work}.

There are some initial efforts addressing Challenges 2 and 3 which we compare in Table \ref{comparisons}. Simulation-based tools \cite{8416816, 9319909, 10.1145/2694344.2694362} enable rapid prototyping but are not substitutes for real hardware execution. Prior work such as \cite{alonso, cong, Baskin_2018, 8425399, 8715174, 9114652, weiwen} proposed CNN/DNN partitioning across FPGAs, but are not generalizable to different workloads. Other works such as \cite{fleming} leverage latency-insensitivity (discussed in Section \ref{inter}) to partition the design across FPGAs but expect the user to provide module-to-FPGA mappings in RTL, and perform simulation-based experiments. If HLS front-ends are combined with automatically partitioned designs, users from varied backgrounds are more likely to adopt FPGAs. Recent virtualization-based work \cite{10.1145/3373376.3378491, 9499729, 10.1145/3445814.3446699} also leverages latency-insensitive design to partition the workload, but virtualizes the FPGA by creating pre-placed and pre-routed static regions to which user logic is mapped. This increases the area overheads and degrades the customizable nature of FPGAs. Most prior works take advantage of the high-bandwidth networking capabilities available in modern FPGAs in the form of Ethernet-compatible ports, but do not formulate their design partitioners in a way that minimizes and hides this latency. We find that none of these prior works consider coupling intelligent floorplanning of the compute modules and interconnect pipelining with HLS compilation. This step is crucial in achieving designs with high frequency as we discuss in Section \ref{background}. Also, none of the prior works consider the topology of the networked-FPGA infrastructure. This might result in the suboptimal mapping of compute modules to devices, and is a major hurdle in the scalability of the tool beyond two FPGAs. We discuss more details of prior works in Section \ref{prior_work}.

To this end, we propose TAPA-CS which takes as input any large-scale dataflow workload expressed in C/C++, and automatically partitions and maps it to a cluster of modern FPGAs during HLS compilation. TAPA-CS couples the process of intra- and inter-FPGA floorplanning with interconnect pipelining to ensure high frequency. TAPA-CS is built upon the latest progress in dataflow-based FPGA HLS design tools \cite{9444053, 10.1145/3609335, 10.1145/3431920.3439289}. Our main contributions are as follows:

\begin{enumerate}
  \item Integrate two layers of floorplanning (inter- and intra-FPGA) and interconnect pipelining with HLS compilation using an Integer Linear Programming (ILP)-based resource allocator which takes into account network topology, internal FPGA chip layout, and the resource requirements of the input workload. This ensures generalizability to different workloads and network topologies, low resource congestion on-board, and high frequency designs.
  \item Utilize the latency-insensitive nature of the dataflow design to partition it across devices, allowing us flexibility in implementing the inter-FPGA communication infrastructure.
  \item Raise the abstraction level of programming cloud-scale FPGAs by hiding the cluster complexity and virtualizing multiple devices as one from the user perspective. 
  \item We test TAPA-CS on different applications ranging from systolic-array based CNNs, stencil designs, Page Rank, and KNN on 2-8 FPGAs. Across the tested designs we achieve an average throughput increase of 2.1$\times$, 3.2$\times$, and 4.4$\times$ using 2, 3, and 4 FPGAs. We also improve the design frequency between 11-116\% compared with Vitis HLS.
  \item Discuss the computation to communication trade-offs of each benchmark and how they impact the scalability to multiple nodes and FPGAs. 
\end{enumerate}

\section{Background}\label{background}
\begin{figure}
  \includegraphics[width=\columnwidth]{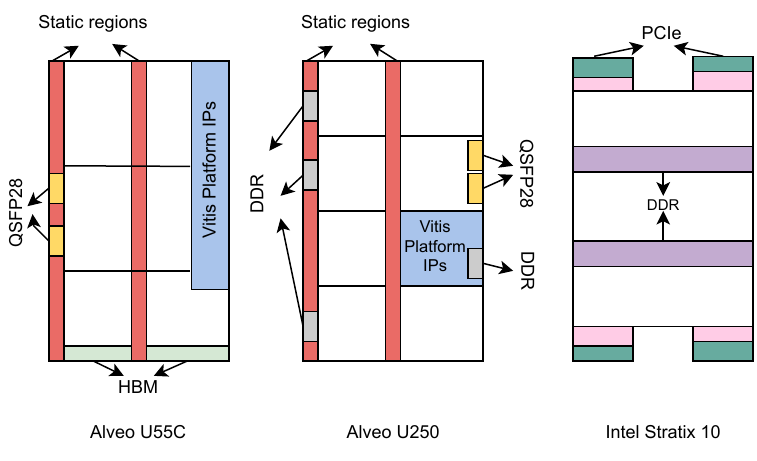}
  \caption{Architecture examples of modern FPGAs.}
  \label{fig:2}
\end{figure}

FPGAs consist of programmable logic organized in the form of a 2-D grid. This programmable region consists of look-up tables (LUTs) which can implement the truth table for any 6-input function. In recent years, FPGAs have also come to include several hard IPs such as PCIe IPs, DDR/HBM controllers, and other platform-specific IPs between the programmable logic regions. While these IPs improve design performance, they have a fixed location on-board and consume a significant amount of logic around them. AMD/Xilinx UltraScale+ FPGAs \cite{ultrascale} are also organized into multiple dies separated by silicon interposers. Crossing these die boundaries results in a much higher delay than on-die interconnect. Similar trends are also observed in case of Intel FPGAs \cite{stratix-10}. Both AMD and Intel FPGA boards expose physical interfaces in the form of PCIe ports and networking interfaces in the form of Ethernet-compatible QSFP28 ports. Figure \ref{fig:2} describes the chip layout of the Alveo U55C, U250 cards and the Intel Stratix 10 cards. 


\begin{figure}
  \includegraphics[width=\columnwidth, keepaspectratio]{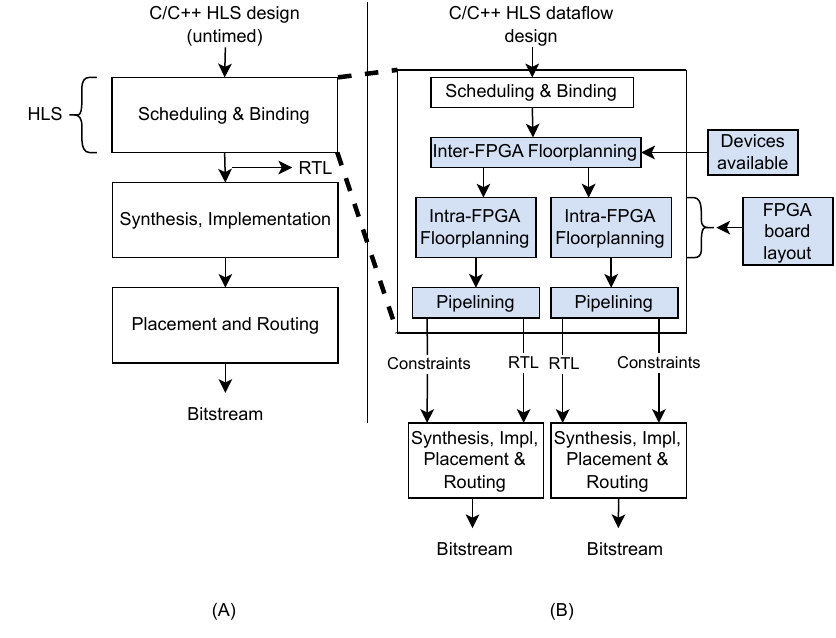}
  \caption{(A) Typical FPGA compilation flow, (B) Additions by TAPA-CS are highlighted in blue.}
  \label{fig:3}
\end{figure}
Accelerators for such FPGAs can be designed using commercial tools like Vitis \cite{vitis_hls} and Intel HLS \cite{intel_hls}. Figure \ref{fig:3} (A) depicts the key steps involved in such toolflows. First, the untimed C/C++ input is converted into a timed RTL using HLS. Next, the timed RTL is passed for synthesis, placement and routing where the logic is mapped to the physical hardware. While these tools simplify the design process, they often suffer from poor quality of results compared with an expert-tuned RTL. The key reason behind this is that HLS tools cannot correctly estimate the final placement of compute modules on the board, and insert insufficient number of clock boundaries (registers) between them while converting the untimed input into a timed output. Due to this, several connections remain underpipelined, degrading the final frequency. Therefore, it is important to provide the tool a global view of the chip layout and the placement of the compute modules \textit{during} HLS compilation. 

Prior work such as Autobridge \cite{10.1145/3431920.3439289} integrates a coarse-grained floorplanning step with interconnect pipelining in HLS compilation for optimization on a single FPGA device. Other approaches such as \cite{10.1145/2554688.2554775, 10.1145/268424.268425} also attempt to provide HLS with a layout of the device, but either suffer from high runtime or evaluate their method only on small applications. TAPA \cite{chi2021extending} extends HLS to feature fast compilation, and an expressive programming model for task-parallel programs. It exposes user-friendly APIs which decouple communication and computation, allowing the user great flexibility in implementing the inter-module communication patterns.The new version of TAPA \cite{10.1145/3609335} integrates AutoBridge to improve design frequency, and we use this as the single FPGA baseline to evaluate TAPA-CS in Section \ref{eval}.
\section{Motivating Example of TAPA-CS}\label{motivating-example}
We demonstrate the importance of using TAPA-CS through the KNN application. Through this example, we aim to dispel the common misconception that scale-out acceleration is only useful when the design cannot be routed on a single FPGA. We find that even when designs can be successfully routed on a single device, span-out acceleration across multiple devices allows the design to efficiently utilize on-chip memory and HBM. 
\par
\begin{figure}
  \centering
  \includegraphics[height=6cm, keepaspectratio]{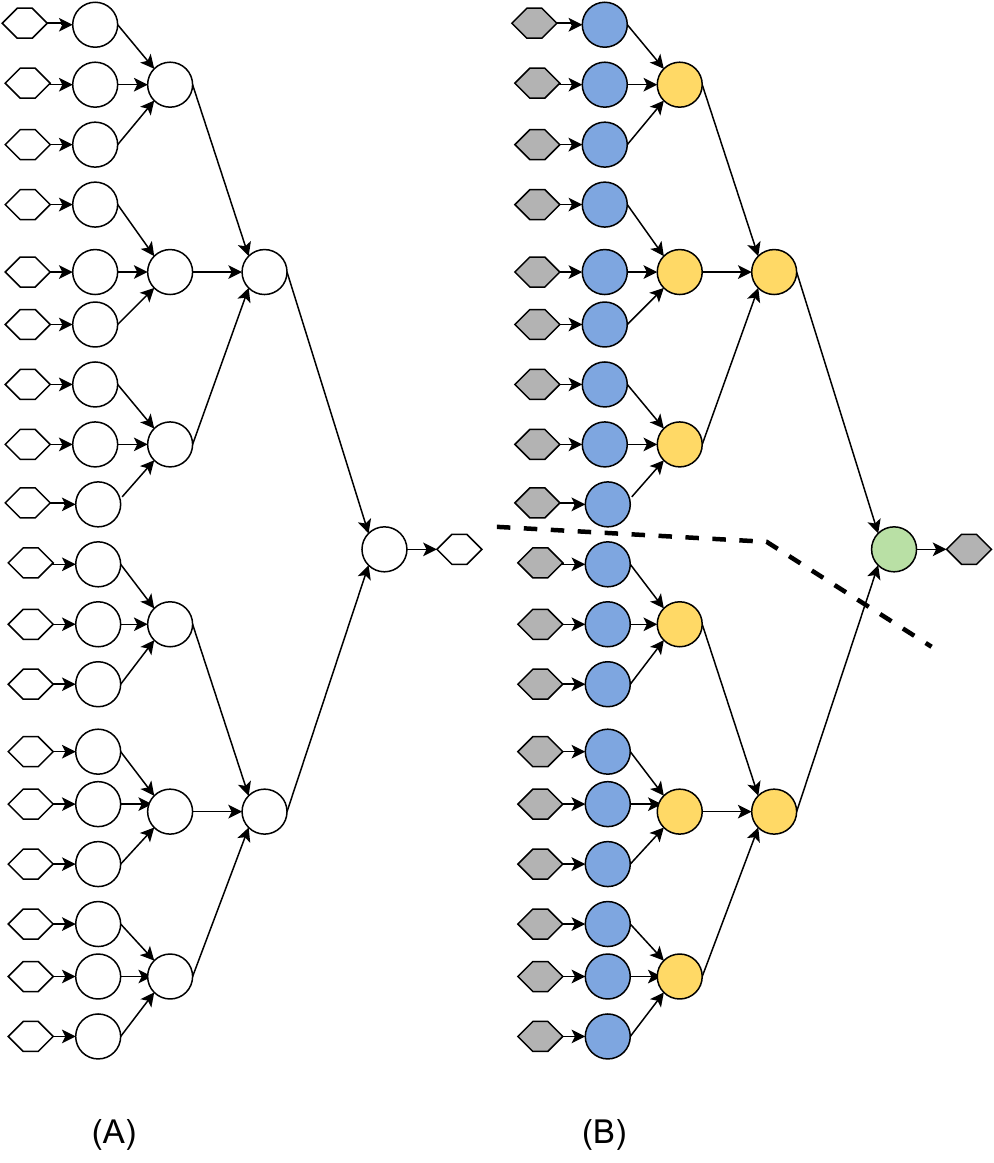}
  \caption{(A) Topology of the KNN application as found by the graph extraction step of TAPA-CS. Here, circles are compute modules (explained in Section \ref{motivating-example}), and hexagons indicate HBM access. (B) Partition found by TAPA-CS is indicated by the dashed line. }
  \label{fig:partitionexample}
\end{figure}

 We use the KNN algorithm presented in \cite{9415564}. The topology of the application is as illustrated in Figure \ref{fig:partitionexample} (A). There are two major phases in this algorithm. The first phase calculates the distance of an input query data point with every other datapoint in the dataset. This phase is indicated by the blue modules in Figure \ref{fig:partitionexample} (B). Given that the dataset contains N data points, each represented as a D-dimensional feature vector, this phase has a computational and memory access complexity of $O(N*D)$ for a single query. Since the KNN application is commonly used on very large datasets with large features, the computational and memory access cost quickly scales up. The second phase sorts the N distances calculated in Phase 1 and returns the top K nearest neighbors. This phase is indicated by the yellow modules in Figure \ref{fig:partitionexample} (B). Since K is usually small and we only need to sort for the K smallest distances, the complexity of this phase is $O(N*K)$. Lastly, the green module in Figure \ref{fig:partitionexample} accumulates the final result and stores it back in HBM. 

\par The design generated by traditional CAD tools results in a low design frequency of 165MHz and high latency. There are three main reasons for this. First, this design utilizes a port width and buffer size of 256 bits and 32KB which only saturates about 51.2\% of the per-bank HBM bandwidth. Prior work has also found that in case of memory-bound applications when multiple processing elements (PEs) access the HBM channels, the achievable HBM bandwidth can drop to as low as 9.4GBps \cite{10.1145/3431920.3439301}. We find that the optimal port width and buffer size which allows us to saturate the per-bank bandwidth is 512 bits and 128KB respectively. This configuration however, results in very high resource utilization in the lower die, leading to a failure in the routing phase of traditional CAD tools. Second, since a smaller buffer size is used, there is a higher number of HBM accesses. Given that HBM accesses are about 76x slower than on-chip memory access, an efficient design should use on-chip memory as much as possible. Third, this design does not feature any intelligent floorplanning or interconnect pipelining. Even with intelligent floorplanning and interconnect pipelining on a single FPGA, the design frequency increases to 198MHz and still incurs a high latency since the HBM bandwidth cannot be saturated. 
\par TAPA-CS includes all these considerations to generate an optimized KNN implementation automatically partitioned across two FPGAs as shown in Figure \ref{fig:partitionexample} (B). This provides sufficient resources in the lower die to route the optimal port width and buffer size configuration. Also, since the input data is divided between the two FPGAs, the compute load is balanced and neither FPGA is idle. The designs generated by TAPA-CS result in a design frequency of 300MHz and are 2.0x faster than designs on a single FPGA. Therefore, it is often a misconception that a multi-FPGA design is worse than a single FPGA design if it could be routed successfully on a single FPGA. Using multiple FPGAs can expose higher HBM bandwidth per compute module (aiding the performance of memory-bound applications), and enable the successful routing of larger designs (aiding the design of compute-bound applications). In Section \ref{eval}, we scale this KNN design to 2-8 FPGAs by increasing the number of PEs. 

\section{TAPA-CS Design}\label{core-design}
\begin{figure*}[htbp]
\centering
  \includegraphics[width=14cm, keepaspectratio]{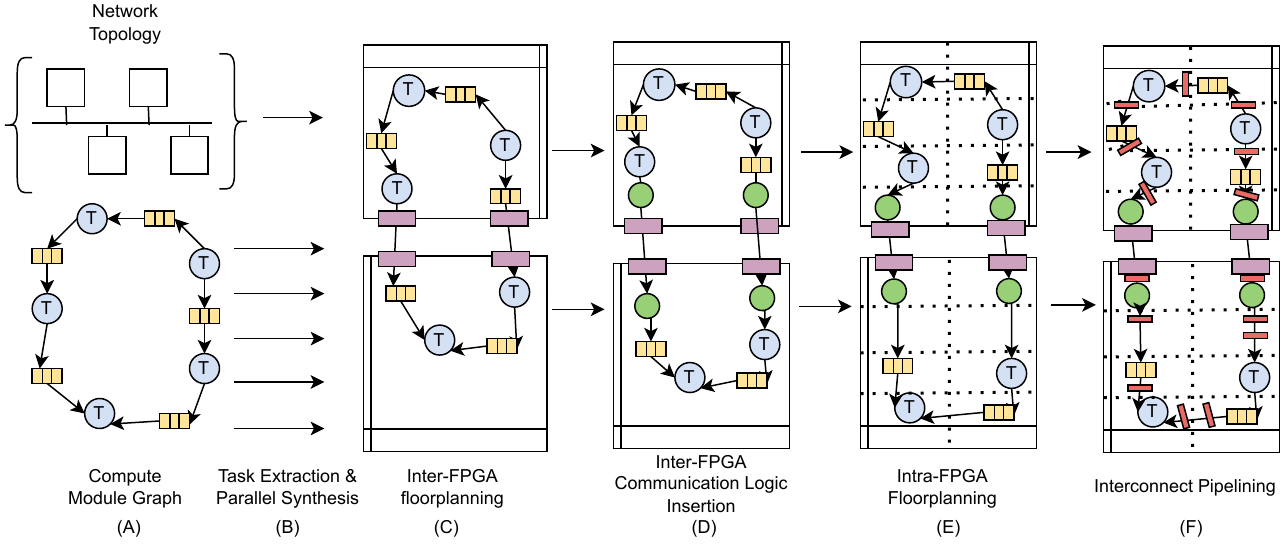}
  \caption{Key Steps in TAPA-CS}
  \label{TAPA-CS-fig}
\end{figure*}

\subsection{Problem Formulation}
TAPA-CS takes as input a C/C++ dataflow program currently written in the TAPA-format \cite{chi2021extending} in which each function compiles into an RTL module and communicates with other functions using FIFOs. We also take the network topology and number of FPGAs present in the user cluster as input. 

We model the input program as a graph \(G(V, E)\), where each vertex ($v_i \in V$) is one of the functions, and the edges ($e_i \in E$) correspond to the FIFOs connecting them as shown in Figure \ref{TAPA-CS-fig} (A). Our goal is to map each vertex $v_i$ to an FPGA $F_i$ in the cluster such that the inter-FPGA communication cost is minimized while ensuring the compute-load between the multiple FPGAs is balanced. We explain how we model this cost in Section \ref{inter}. We also apply several chip-level optimizations to ensure high frequency as discussed in Section \ref{intra}. 


\subsection{Key Steps}
There are seven major steps in TAPA-CS:
\begin{enumerate}
  \item \textbf{Task graph construction}: We model the input workload as a graph $G(V, E)$ where nodes are compute modules and edges are the FIFOs connecting them as shown in Figure \ref{TAPA-CS-fig}(A).
  \item \textbf{Task extraction and parallel synthesis}: We extract and synthesize each compute module in parallel providing HLS an accurate resource utilization profile as shown in Figure \ref{TAPA-CS-fig} (B).
  \item \textbf{Inter-FPGA floorplanning}: We use the resource utilization profile to intelligently floorplan this design across multiple FPGAs connected to each other through any topology (daisy-chained, ring, bus, star, mesh, hypercube, etc.) and data transfer protocol (PCIe, Ethernet, etc.). This step allows us to address the key limitation of traditional CAD toolflows (discussed in Section \ref{background}) by providing the scheduling and binding stage an accurate view of the topology and available programmable resources. Our goal is to assign each compute module to an FPGA as shown in Figure \ref{TAPA-CS-fig} (C). We explain the details of this step in Section \ref{inter}.
  \item \textbf{Inter-FPGA communication logic insertion}: After mapping the design to multiple devices, we add the inter-FPGA communication logic as shown in Figure \ref{TAPA-CS-fig}(D). We discuss details of the communication logic in Section \ref{networking-interface}.
  \item \textbf{Intra-FPGA floorplanning}: We intelligently floorplan the design across each FPGA chip by providing the scheduling and binding stage information about the locations of the hard IPs, and I/O ports, and the internal chip layout as shown in Figure \ref{TAPA-CS-fig} (E). Section \ref{intra} details how we formalize this information and divide the FPGA into multiple slots.
  \item \textbf{Interconnect Pipelining}: We add pipeline registers to the interconnect at the slot crossings to ensure high frequency designs as shown in Figure \ref{TAPA-CS-fig} (F). We also ensure correctness and that the final design execution cycles are not compromised by this interconnect pipelining step as discussed in Section \ref{pipelining}.
  \item \textbf{Bitstream generation}: Finally, the optimized designs and floorplanning constraints found by TAPA-CS are passed back into the traditional CAD stack to produce the bitstreams. 
\end{enumerate}

TAPA-CS is completely integrated with existing FPGA CAD toolflows. While the method proposed in TAPA-CS can be applied to any multi-FPGA setup, for the scope of this paper, we test TAPA-CS on Xilinx/AMD Alveo boards. In the following Sections, we discuss each of these steps in detail and display the features through which our tool achieves high throughput and frequency accelerators.

\subsection{Inter-FPGA Module Mapping}\label{inter}
\begin{figure}
\centering
  \includegraphics[height=5cm, keepaspectratio]{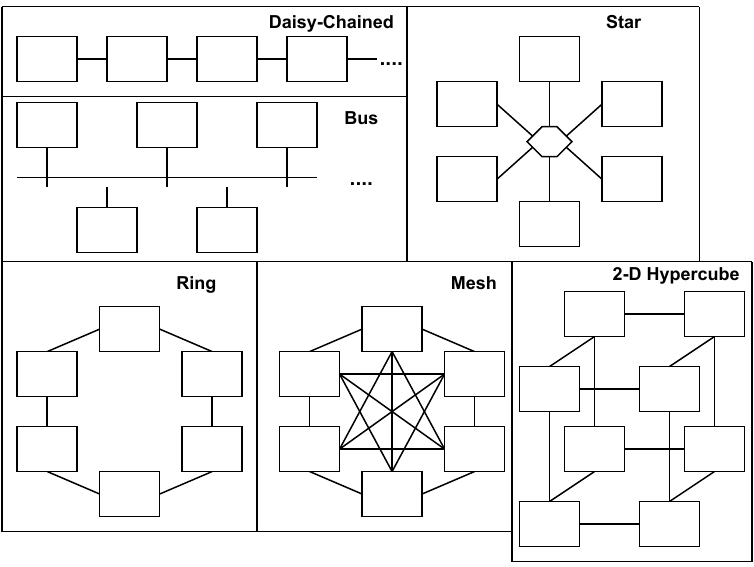}
  \caption{Network Topologies}
  \label{topo-devices}
\end{figure}

In this step, we provide the traditional CAD toolflows with a view of the available FPGA devices, their topologies, and the hierarchy of the user application, so that we can automatically find the optimal module-to-FPGA mapping. Design partitioning is enabled by the latency-insensitive nature of dataflow designs. Latency-insensitive design \cite{945302, 810667} decouples the design of the interconnect from that of the compute modules. This allows the designer to add interconnect pipeline registers, and connections over different networks with arbitrarily long latency between compute modules, without affecting the functional correctness of the design. 

In this mapping step, our goal is to minimize the high inter-FPGA communication cost while ensuring there is no resource congestion. If we are using two FPGAs, we consider the total available region to be divided into two grids as shown in Figure \ref{TAPA-CS-fig} (C). 

Now, the placement task reduces to assigning each task to one of the two grids. For this, we use an ILP-based formulation to obtain exact partitioning solutions. While heuristic solvers are faster, ILP allows an accurate solution. We also show in Section \ref{eval} that both our intra- and inter- FPGA partitioning algorithms only add between 2.8-49.7 seconds overhead to the overall compilation flow for number of compute modules ranging from 30 to 493, making the method scalable. 

Let the binary variable denoting whether vertex $v_i$ is placed on device $F_i$ or not be $v_d$. Let the task \(v_i\in V\) have a resource utilization profile of \(v_{area}\). If \(r_v\) is the resources used by the set of tasks already placed in device $F_i$, then, before placing a new task \(v_i\) in this device, we need to ensure that there are enough resources. That is, for each type of on-chip resource, 

\begin{equation}
  \sum_{v\in r_v}v_d\times v_{area} < T,
\end{equation}
where T is the threshold of utilization for each resource. Next, our ILP solver ensures that the cost of inter-FPGA communication is minimized. Therefore, we consider the placement of all the neighbors of task $v_i$ in the cost function as follows:
\begin{equation}
  \sum_{e_{ij} \in E}{e_{ij}}.width\times dist(F_i, F_j) \times \lambda
\end{equation}
where $e_{ij}.width$ is the bitwidth of the FIFO channel connecting the two vertices $v_i$ and $v_j$, function $dist(F_i, F_j)$ is a metric of the cost of communication between tasks $v_i$ and $v_j$ placed on the same or different FPGAs, and $\lambda$ is a scaling factor to adjust the cost for different data transfer protocols like Ethernet and PCIe. 

The communication cost function $dist(F_i, F_j)$ depends on the topology of the network-connected FPGAs. Consider the network topologies shown in Figure \ref{topo-devices}. In case the FPGAs are daisy-chained, 
\begin{equation}
dist(F_i, F_j) = |F_i.device\_num - F_j.device\_num|
\end{equation}
where $F_i.device\_num$ and $F_j.device\_num$ are the device IDs associated with the FPGAs. 
Similarly, in the case of a bidirectional ring topology, the distance metric changes to:
\begin{multline*}
dist(F_i, F_j) = \min (|F_i.device\_num - F_j.device\_num|, \\(total\_num - |F_i.device\_num - F_j.device\_num|))
\end{multline*}
where $total\_num$ is the total number of FPGAs in the ring.

The scaling factor $\lambda$ is used to adjust the cost in a system with multiple interconnection media. We use Ethernet-based connections offering 100GBps bandwidths as the baseline and scale the cost for other media accordingly. For example, if the interconnection used is PCIe Gen3x16, then the cost is scaled by a factor of 12.5 compared with the cost of using Ethernet-based connections. 


Note that our partitioner does not always recommend the min-cut. For the placement of each module, we consider the resource and communication cost added by its placement on- and off-chip. In case the module can be accommodated on the same chip as its neighbors, we would pay a high price by moving the compute module off-chip than placing it on-chip. However, if the placement of this module on-chip results in congestion (which in turn lowers design frequency), it would be more beneficial to place the module off-chip even at the cost of increased inter-FPGA connections. This trade-off ensures that we can achieve high frequency designs on both FPGAs. 

\subsection{Inter-FPGA Communication}\label{networking-interface}
\begin{figure}[htbp]
 \centering
 \includegraphics[height=5cm, keepaspectratio]{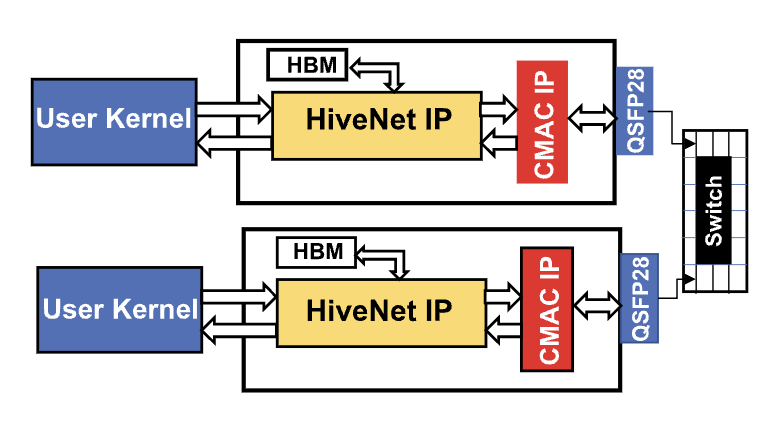}
 \caption{AlveoLink features as described in \cite{alveolink}}
 \label{alvlink_figure}
\end{figure}

 Despite the excellent opportunity to leverage networking capabilities in modern FPGAs, existing CAD tools do not explicitly support networking. TAPA-CS supports a library of inter-FPGA communication protocols, such as Ethernet-based RoCE v2, and PCIe-based P2P DMA \cite{p2p}.  However, for the scope of this paper, we limit our discussions and evaluations to using the QSFP28 Ethernet ports. Here, we use AlveoLink \cite{alveolink} described in Figure \ref{alvlink_figure} to illustrate how to add networking support to the existing toolflows. AlveoLink ensures reliable, lossless, and in-order data transfer with a low resource overhead of $\sim$5\% on the Alveo U55C cards. It offers a low round-trip data transfer latency of 1 $\mu$s between two FPGAs. AlveoLink is 12.5x faster compared with PCIe Gen3x16 -based connections. AlveoLink's main components are as follows:
\begin{enumerate}
  \item HiveNet IP: This is a Vitis-compatible implementation of the RoCE v2 protocol which directly connects to user kernels.
  \item CMAC kernel: This is a board-specific interface between the signals detected at the QSFP28 ports and the signals detected at the commodity network.
\end{enumerate}

\begin{figure}[htbp]
 \centering
 \includegraphics[height=5cm, keepaspectratio]{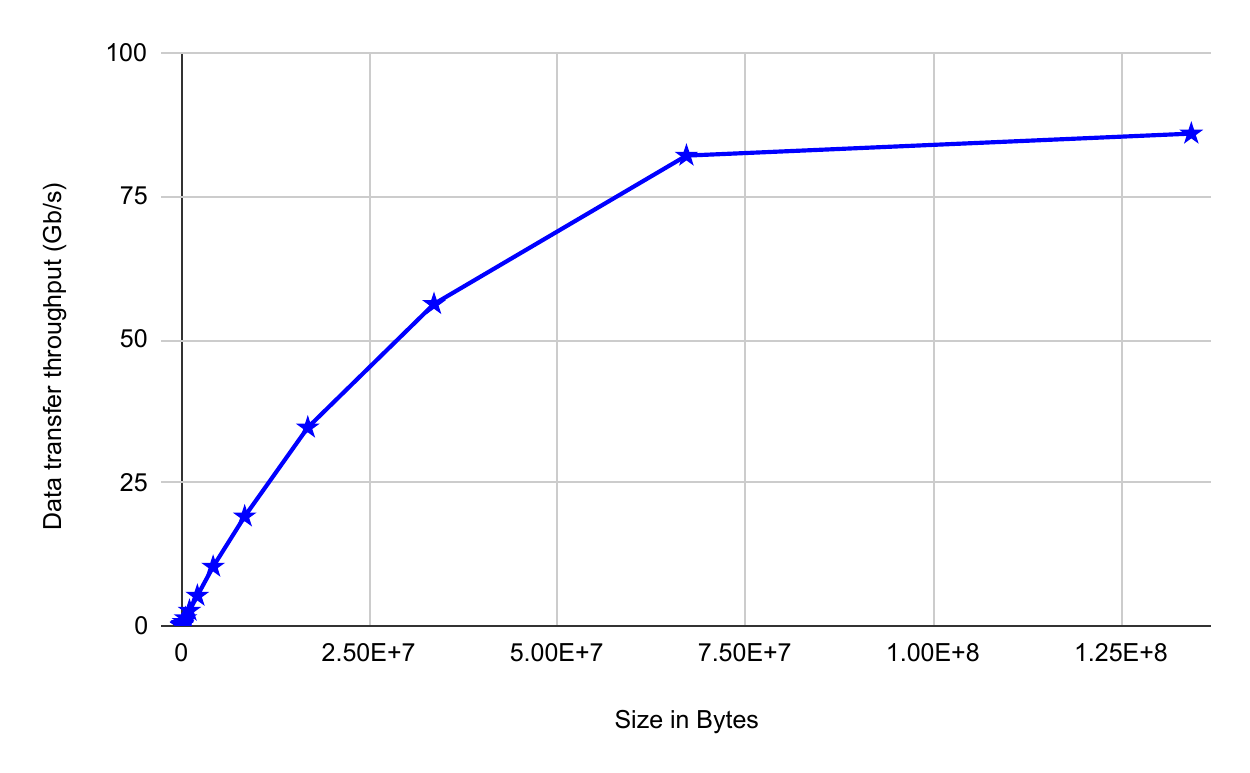}
 \caption{Data transfer throughput (Gbps) achieved by AlveoLink for different transfer sizes.}
 \label{alvlink_figure_throughput}
\end{figure}

Figure \ref{alvlink_figure_throughput} shows the data transfer throughput achieved by AlveoLink for different data transfer volumes. Note that this throughput is per-port per-FPGA. We discuss the overheads added by AlveoLink in Section \ref{overheads_TAPA-CS}.

\subsection{Intra-FPGA Module Mapping} \label{intra}
After assigning each task to the set of devices in our cluster and adding the inter-FPGA communication interfaces, the next step is to apply a similar top-down partitioning approach to each FPGA (Figure \ref{TAPA-CS-fig}(E)). To formalize the device-specific information and present it to the scheduling and binding stage, we view each FPGA as a grid divided into slots by the hard IPs and static regions. For example, the Alveo U55C card shown in Figure \ref{fig:2} is presented to TAPA-CS as a grid with 6 slots divided into two columns and 3 rows. Our goal is to place each vertex in one of these slots based on the resource utilization ratios per slot and the cost of connecting this module to all its neighbors. In this step, our goal is to minimize the cost of inter-die communication. Therefore, the new cost function is:

\begin{equation}
  \sum_{e_{ij} \in E}{e_{ij}}.width\times (|{v_i}.row-{v_j}.row|+|v_i.col-v_j.col|)
\end{equation}

where $v_i.row$, $v_j.row$ represent the rows in which tasks $v_i$ and $v_j$ are placed respectively, and the same for the columns. We continue such a two-way ILP-based partitioning scheme until we divide each FPGA into eight grids. 

One of the key considerations in this step is the optimal usage of the HBM channels. Consider the example of the U55C cards. While the HBM offers a high aggregate bandwidth of 460GBps, it is still 76x slower than on-chip data accesses. Therefore, it is important to utilize the HBM channels exposed to the user kernel optimally. As shown earlier in Figure \ref{fig:2}, all the HBM channels on-board the U55C are exposed in the bottom-most die. Suboptimal HBM channel binding can result in large routing delays and increase congestion in this die, leading to routing failure. Therefore, TAPA-CS supports an automatic HBM channel binding exploration where we find the optimal mappings based on the workload characteristics. 

\subsection{Interconnect Pipelining}\label{pipelining}
Following the intra-FPGA optimizations, we also conservatively pipeline the interconnect at all slot-crossings to prevent long delays from degrading the final clock frequency. In contrast to prior work like \cite{1458143, 945302}, we conservatively pipeline all slot-crossing wires because each of our compute modules compiles into an RTL controlled by a finite state machine (FSM). Therefore, it is difficult to estimate the latency added by the pipelining step. Next, to ensure that the design throughput is not negatively affected by the additional pipeline registers, we also balance the latency of parallel paths based on cut-set pipelining \cite{cut-set} as shown in \cite{10.1145/3431920.3439289}. In this step, the latency added by reconvergent paths is balanced to ensure that final correctness is not impacted. The pipeline FIFOs are indicated in red in Figure \ref{TAPA-CS-fig}(F).


\section{Evaluation} \label{eval}
We implement TAPA-CS in Python and integrate it with Vitis 2022.1. Either MIP \cite{mip} or the Gurobi solver \cite{gurobi} (free for academia) can be used to solve our ILP formulations described in Sections \ref{inter} and \ref{intra}.
We test TAPA-CS on a server equipped with two nodes, each featuring four Xilinx Alveo U55C cards connected through their QSFP28 ports in a ring topology using active cables. The total programmable resources available per card can be found in Table \ref{available}. This board can achieve a maximum design frequency of 300MHz. The server features a 128 core AMD EPYC 7V13 CPU operating at 2.45GHz.
\begin{table}
\centering
\begin{tabular}{|c|c|}
\hline
Resource Type & Available \\
\hline
LUT & 1146240 \\
\hline
FF & 2292480 \\
\hline
BRAM & 1776 \\
\hline
DSP & 8376 \\
\hline
URAM & 960 \\
\hline
\end{tabular}
\caption{Resource availability on the Alveo U55C cards.}
\label{available}
\end{table}

\subsection{Benchmarks and Baselines}
We evaluate TAPA-CS over multiple variations of the following benchmarks:
\begin{enumerate} 
  \item Stencil Dilate: This is a 2-D 13-point stencil kernel from the Rodinia HLS benchmark \cite{8457638, 10.1145/3572547} which is used in biomedical research to track leukocytes in blood vessels. We test this kernel over 64 to 512 iterations.
  \item Page Rank created by \cite{chi2021extending}: This kernel features four PEs and one central controller with dependency cycles between the compute modules. It implements the algorithm described in \cite{Page1999ThePC}. We test this design over multiple networks taken from the SNAP dataset collection \cite{snapnets}.
  \item KNN created by \cite{9415564}: This kernel contains 17 compute modules implementing an optimized accelerator for calculating each data point's distance to its neighbor, and sorting the distances to obtain the K-nearest neighbors. We test this design across varying input sizes and feature dimensions.
  \item Systolic-array CNN accelerators created by AutoSA \cite{10.1145/3431920.3439292}. This systolic array accelerator consists of multiple PEs arranged in a grid format, with a total of 493 compute modules. The CNN we choose is an implementation of the third layer of the VGG model \cite{simonyan2015deep}. We test TAPA-CS on multiple grid dimensions ranging from 13 x 4 to 13 x 20.
\end{enumerate}

\begin{figure}
  \includegraphics[width=\columnwidth]{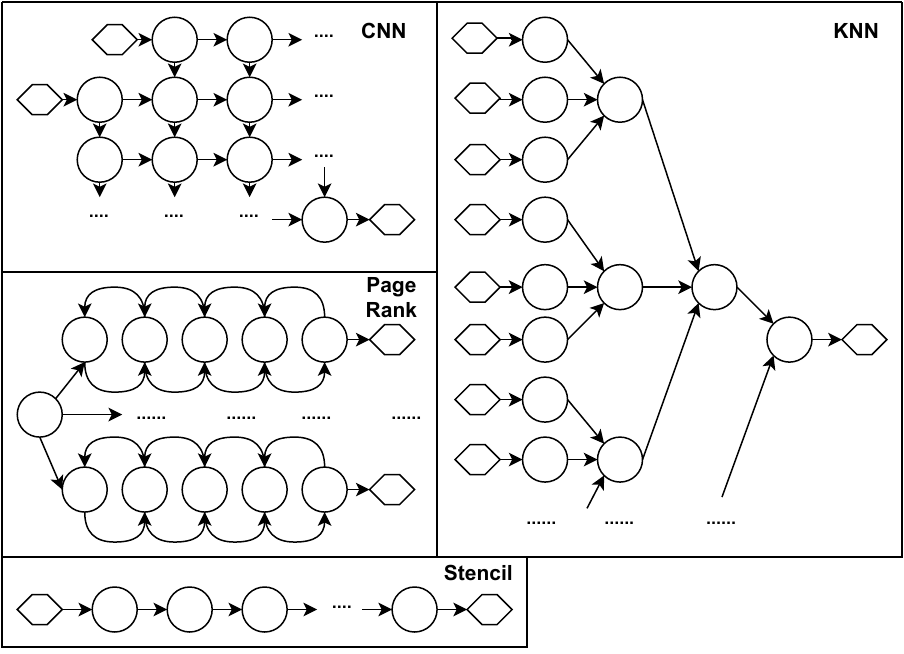}
  \caption{Topology of benchmarks. Here, circles represent compute modules while hexagons represent HBM access.}
  \label{topo}
\end{figure}

Figure \ref{topo} depicts the topology of each of the benchmarks. We compare the latency and frequency of TAPA-CS with two baselines, namely F1-V and F1-T. Here, F1-V is a single FPGA implementation generated through Vitis HLS and F1-T is a single FPGA implementation generated through TAPA/AutoBridge \cite{10.1145/3609335}. The TAPA-CS designs are denoted by F2 (2 FPGAs), F3 (3 FPGAs), and F4 (4 FPGAs). 

Table \ref{tab:summary} summarizes the speed-up obtained by TAPA-CS for each benchmark averaged across all the datasets and configurations tested. Note that the speed-ups are normalized with respect to the single FPGA design generated by Vitis HLS. We discuss more details of each application in the following Sections and also discuss the reasons for the varying speed-ups. 

We discuss the floorplanning and resource overheads added by TAPA-CS in Section \ref{overheads_TAPA-CS}. Lastly, we evaluate designs spanning across multiple nodes and 8 FPGAs in Section \ref{scalability}.

\begin{table}[h]
    \centering
    \begin{tabular}{|c|c|c|c|c|c|}
    \hline
        Benchmark & F1-V & F1-T & \textbf{F2} & \textbf{F3} & \textbf{F4} \\
        \hline
        Stencil & 1 & 1.25$\times$ &\textbf{1.71$\times$} & \textbf{2.37$\times$} & \textbf{3.06$\times$} \\
        \hline 
        PageRank & 1 & 1.54$\times$ & \textbf{2.64$\times$} & \textbf{4.28$\times$} & \textbf{5.98$\times$} \\
        \hline 
        KNN & 1 & 1.2$\times$ & \textbf{1.72$\times$} & \textbf{2.53$\times$} & \textbf{3.60$\times$} \\
        \hline
        CNN & 1 & 1.1$\times$ & \textbf{1.41$\times$} & \textbf{2.0$\times$} & \textbf{2.54$\times$} \\
         \hline
    \end{tabular}
    \caption{Speed-up of TAPA (F1-T), and TAPA-CS (F2, F3, and F4) normalized against the Vitis HLS (F1-V) baseline. }
    \label{tab:summary}
\end{table}



\subsection{Stencil}\label{stencil-design}
Stencil kernels apply a sliding window (or stencil) of computation over an input array to produce an output array. Such kernels can either be memory-bound or compute-bound based on the input size and the number of iterations. Prior work \cite{10.1145/3572547} found that for a fixed input size, stencil designs with a smaller number of iterations are memory-bound while designs with a larger number of iterations are compute-bound. We chose the Dilate kernel from the Rodinia HLS benchmark to test TAPA-CS. It is a 2D 13-point kernel which we test for an input size of 4096$\times$4096 and iterations varying between 64 and 512. 
\begin{table}
\centering
\begin{tabular}{|c|c|c|}
\hline
Iters & Ops/Byte & Volume (MB)\\
\hline
64 & 208 & 144.22\\
128 & 416& 288.43\\
256 &832 & 576.86\\
512 & 1664& 1153.73\\
\hline
\end{tabular}
\caption{Stencil: Compute intensity (operations / byte of external memory access) and total inter-FPGA data transfer over varying iterations and a fixed input size of 4096$\times$4096.}
\label{stencil-comp-inte}
\end{table}

\begin{figure}
  \centering
  \includegraphics[height=5cm, keepaspectratio]{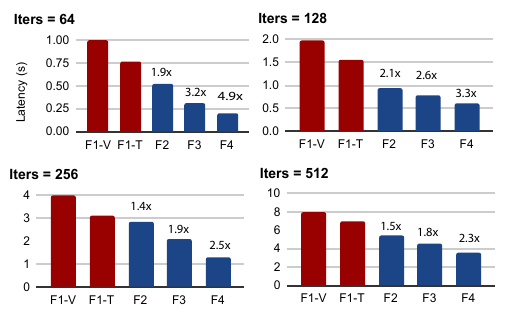}
  \caption{Stencil: Latency comparison between TAPA-CS (F2, F3, F4), TAPA (F1-T), Vitis-HLS (F1-V).}
  \label{fig:stencil-latency-new}
\end{figure}

The single FPGA baseline design can successfully route a design with 64 iterations, 15 PEs, and an HBM access bit-width of 128. A single device cannot support a larger number of PEs, iterations, and bit-widths. We measure the compute intensity (in terms of operations per byte of external memory access) and total inter-FPGA data transfer volume of various iteration configurations in Table \ref{stencil-comp-inte}. Note that the compute intensity calculations assume optimal data reuse. We find that iterations 64 and 128 are memory-bound while iterations 256 and 512 are compute-bound. Therefore, we scale the design to suit the additional on-chip resources and HBM bandwidth available in the multi-FPGA scenario as follows:
\begin{enumerate}
\item Design with 64 and 128 iterations: Increase HBM access bit-width from 128 to 512 as well as the channels used by the design from 32 (single FPGA) to 64 (2 FPGAs), 96 (3 FPGAs), and 128 (4 FPGAs). 
\item Design with 256 and 512 iterations: Increase the number of PEs from 15 (single FPGA), to 30 (2 FPGAs), 60 (3 FPGAs), and 90 (4 FPGAs) keeping the HBM access bit-width at 128.
\end{enumerate}
The overall runtime of the different configurations tested on up to 4 FPGAs is shown in Figure \ref{fig:stencil-latency-new}. For the design with 64 iterations, the 4 FPGA design is 4.9$\times$ faster than the single FPGA design produced through Vitis. The main reasons for this are the increased HBM channel usage and the higher bit-width. However as the number of iterations increases, the relative gains obtained by using multiple FPGAs decreases. The 4 FPGA design with 512 iterations is only 2.3$\times$ faster than the Vitis baseline. This is because in the case of the 4 FPGA design with 64 iterations, each FPGA has a compute-intensity of 52 ops/byte with an inter-FPGA transfer size of 144.22MB. However, for the design with 512 iterations, each FPGA has a compute intensity of 416 ops/byte with an inter-FPGA transfer size of 1153.73MB. Large inter-FPGA transfer sizes lead to higher idle PE time. Also, unlike other applications we discuss in Sections \ref{eval:pagerank} and \ref{eval-knn} (ie., PageRank and KNN), in the case of the stencil application, each FPGA operates sequentially due to the topology of the design. That is, FPGA 2, 3, and 4 lie idle while their predecessor executes. However, as the number of iterations increase and the inter-FPGA transfer sizes rise, this becomes a limitation. 

\begin{figure}
  \centering
  \includegraphics[height=5cm, keepaspectratio]{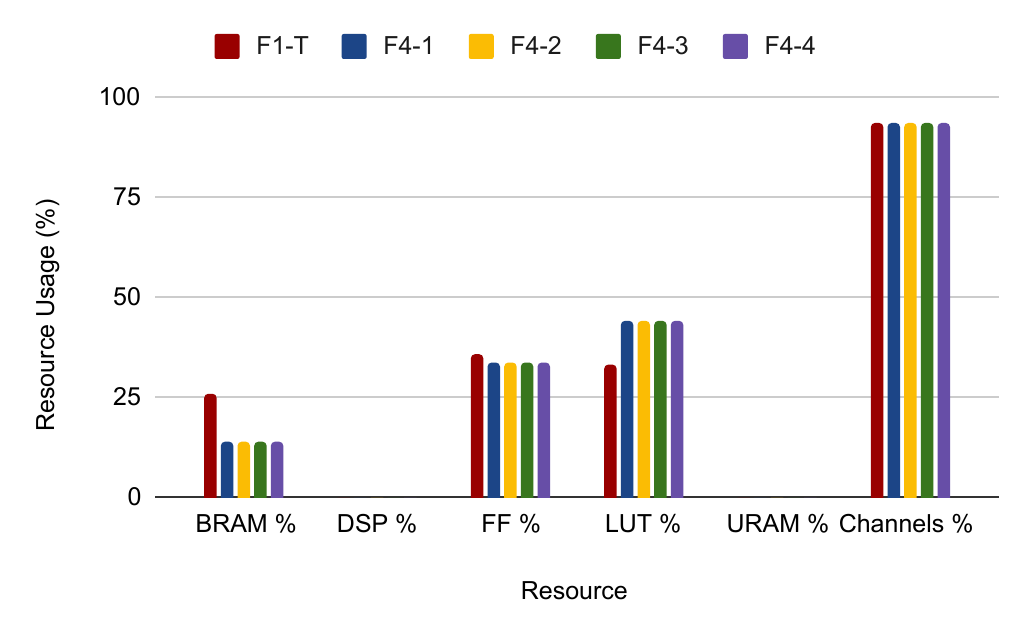}
  \caption{Stencil: Resource utilization of single FPGA baseline (F1-T) and 4-FPGA design (F4). Here, F4-1 to F4-4 denote the 4 FPGAs used in design F4.}
  \label{fig:stencil-resource-new}
\end{figure}

Figure \ref{fig:stencil-resource-new} displays the resource utilization of the single FPGA baseline and each of the 4 FPGAs used in design F4. Since the resource utilization profiles per FPGA in case of F2 and F3 are similar to that of F4, we do not report them. 

The single FPGA baseline achieves a design frequency of 165 MHz using Vitis, 250MHz using TAPA, and each of the 2, 3, and 4 FPGA designs generated by TAPA-CS achieve 300MHz, displaying a 81.8\% increase compared to Vitis, and a 20\% increase compared with TAPA.

\subsection{Page Rank}\label{eval:pagerank}
The topology of the Page Rank application is illustrated in Figure \ref{topo}. This accelerator implements the citation ranking algorithm described in \cite{Page1999ThePC, chi2021extending}. First the input graph is pre-processed on the host and loaded onto the device HBM. Then, the edges are streamed to each PE on-chip, which calculate and propagate weighted rankings from source vertex to destination vertex. These updates are stored back into the HBM before they are accumulated over each vertex to calculate the final ranking. The algorithm runs until convergence and uses an edge-centric model such that edges are traversed to avoid bank conflicts, and not by either source or destination vertices.

The single FPGA baseline can successfully route a design with 4 PEs, using 27 HBM channels to read the edges, and store the intermediate data. We scale the design to increase the number of PEs from 4 (single FPGA), 8 (2 FPGAs), 12 (3 FPGAs), and to 16 (4 FPGAs). Note that in the case of the PageRank application, the inter-FPGA transfer volumes depend on the dataset used and do not change with the number of PEs in the design. Therefore, as the compute intensity of the application scales with the increase in PEs, the inter-FPGA transfer size remains constant for a given dataset. This is in contrast with the Stencil application discussed earlier where the transfer volumes depend on the iterations and not the input size. Also, based on the topology depicted in Figure \ref{topo}, once the FPGA containing the leftmost module (FPGA 1) routing data from the HBM to the different PEs is executed, other FPGAs (FPGA 2-4) can be launched in parallel. These two factors aid in the scalability of the PageRank application. 

We test the design on five networks of varying number of nodes and edges taken from the SNAP dataset \cite{snapnets} shown in Table \ref{tab:datasets-pagerank}. 

\begin{figure}
  \centering
  \includegraphics[height=5cm, keepaspectratio]{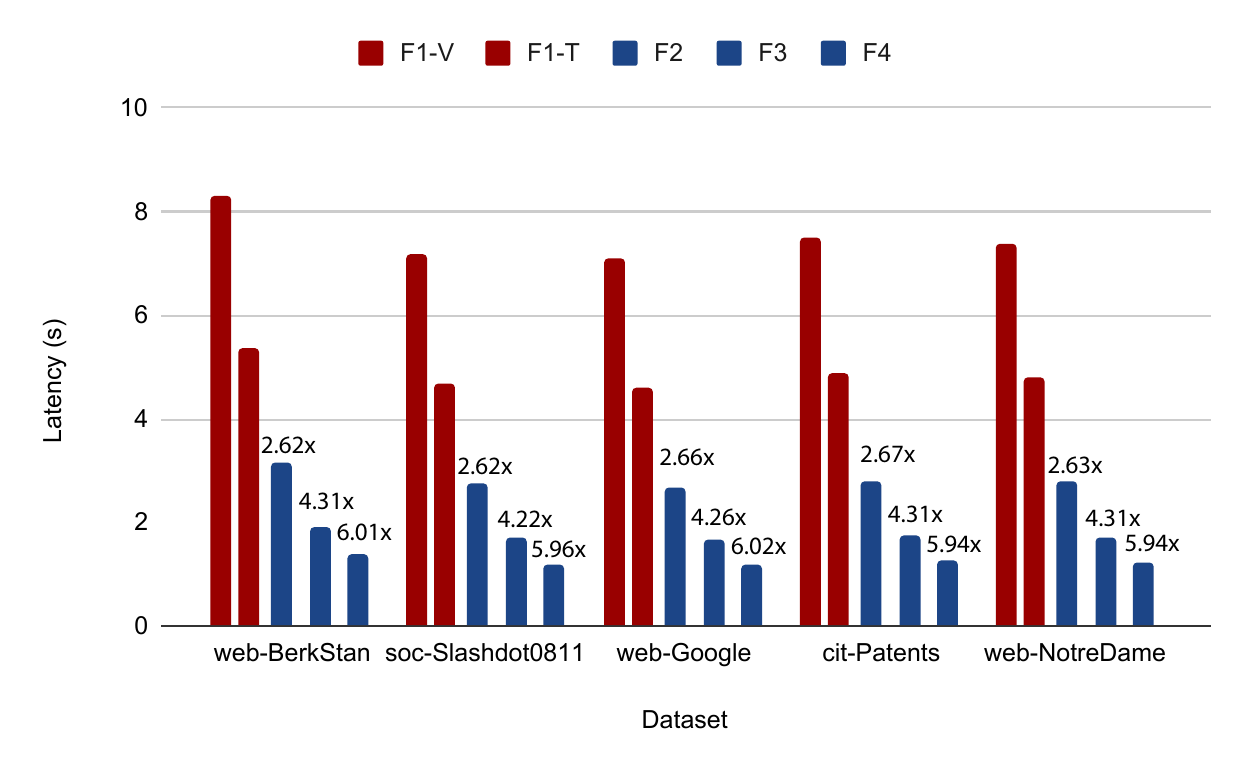}
  \caption{PageRank: Latency comparison between TAPA-CS (F2, F3, F4), TAPA (F1-T), Vitis-HLS (F1-V).}
  \label{fig:pagerank-latency-new}
\end{figure}

The overall runtime of the different PE configurations tested on up to 4 FPGAs is shown in Figure \ref{fig:pagerank-latency-new}. Considering that the data transfer volumes do not increase with the increase in compute intensity for this application, each of the tested configurations benefit from scaling, leading to superlinear speed-ups. The 2, 3, and 4 FPGA designs are on average 2.64$\times$, 4.28$\times$, and 5.97$\times$ faster than the single FPGA Vitis baseline across all the tested datasets. 

\begin{table}[h]
    \centering
    \begin{tabular}{|c|c|c|}
    \hline
    Network & Nodes & Edges \\
    \hline
    web-BerkStan & 685,230 & 7,600,595 \\
    \hline
    soc-Slashdot0811 & 77,360 & 905,468 \\
    \hline 
    web-Google & 875,713 & 5,105,039 \\
    \hline 
    cit-Patents & 3,774,768 & 16,518,948 \\
    \hline 
    web-NotreDame & 325,729 & 1,497,134 \\
    \hline
    \end{tabular}
    \caption{Networks used to test PageRank.}
    \label{tab:datasets-pagerank}
\end{table}

\begin{figure}
  \centering
  \includegraphics[height=5cm, keepaspectratio]{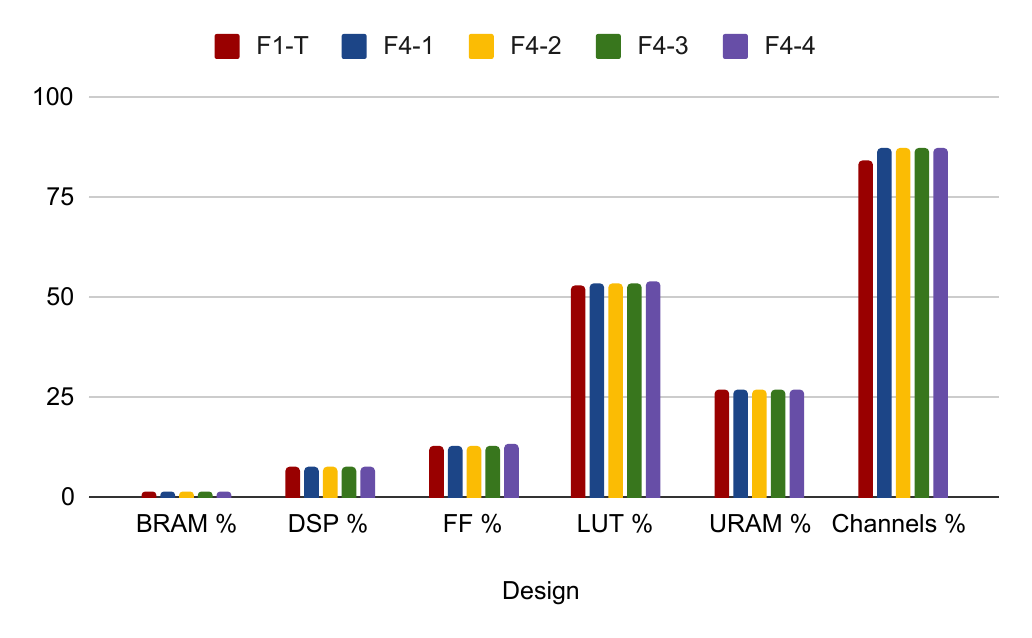}
  \caption{PageRank: Resource utilization of the single FPGA baseline (F1-T) and 4-FPGA design (F4). Here, F4-1 to F4-4 denote the 4 FPGAs used in design F4.}
  \label{fig:pagerank-resource-new}
\end{figure}

Figure \ref{fig:pagerank-resource-new} displays the resource utilization of the single FPGA baseline and each of the 4 FPGAs used in design F4. The resource utilization profiles per FPGA in case of F2 and F3 are similar to that of F4, therefore, we do not report them. 

The single FPGA baseline achieves a design frequency of 123MHz using Vitis, 190MHz using TAPA, and each of the 2, 3, and 4 FPGA designs generated by TAPA-CS achieve 266 MHz, displaying a 116.26\% increase compared to Vitis, and 40\% increase compared with TAPA.

\subsection{KNN}\label{eval-knn}
We use the KNN accelerator designed by \cite{9415564}, which demonstrates quadratic growth in memory access and computational complexity as discussed in Section \ref{motivating-example}. The single FPGA baseline can successfully route a design with a port width and buffer size of 256 bits and 32KB with 27 compute modules (Figure \ref{fig:partitionexample}). Note that the scale of the design is limited by the HBM ports available to the blue compute modules for reading the input data and calculating the pairwise distances. Therefore, we scale the design such that the 2-4 FPGA cases use 36, 54, and 72 blue modules reading from the HBM. The rest of the modules (yellow and green) responsible for sorting the distances and accumulating the final result are also scaled accordingly. To evaluate the effectiveness of the application partitions found by TAPA-CS, we test across varying dataset sizes ($N$) and dimensions ($D$) as shown in Table \ref{knn-params}. The total size of the search space ($N*D*sizeof(float)$) varies from 8MB (N=1M, D=2) to 4GB (N=8M, D=128). 

Note that based on the partition found in Figure \ref{fig:partitionexample}, the size of the inter-FPGA data transfer only depends on the value of K, and is independent over the varying search space. This is because the yellow modules calculate the top-K distances based on the pairwise distances that are computed by the blue modules. Therefore, unlike in the case of the Stencil application discussed earlier, as the computational and memory access complexity of the KNN application scales, the size of the data transfer between FPGAs remains constant. Another difference is that in the case of the KNN application, all FPGAs except the last FPGA (responsible for aggregating the final data using the green module), can be launched completely independently, and do not require any data from each other.

\begin{figure}
  \centering
  \includegraphics[height=5cm, keepaspectratio]{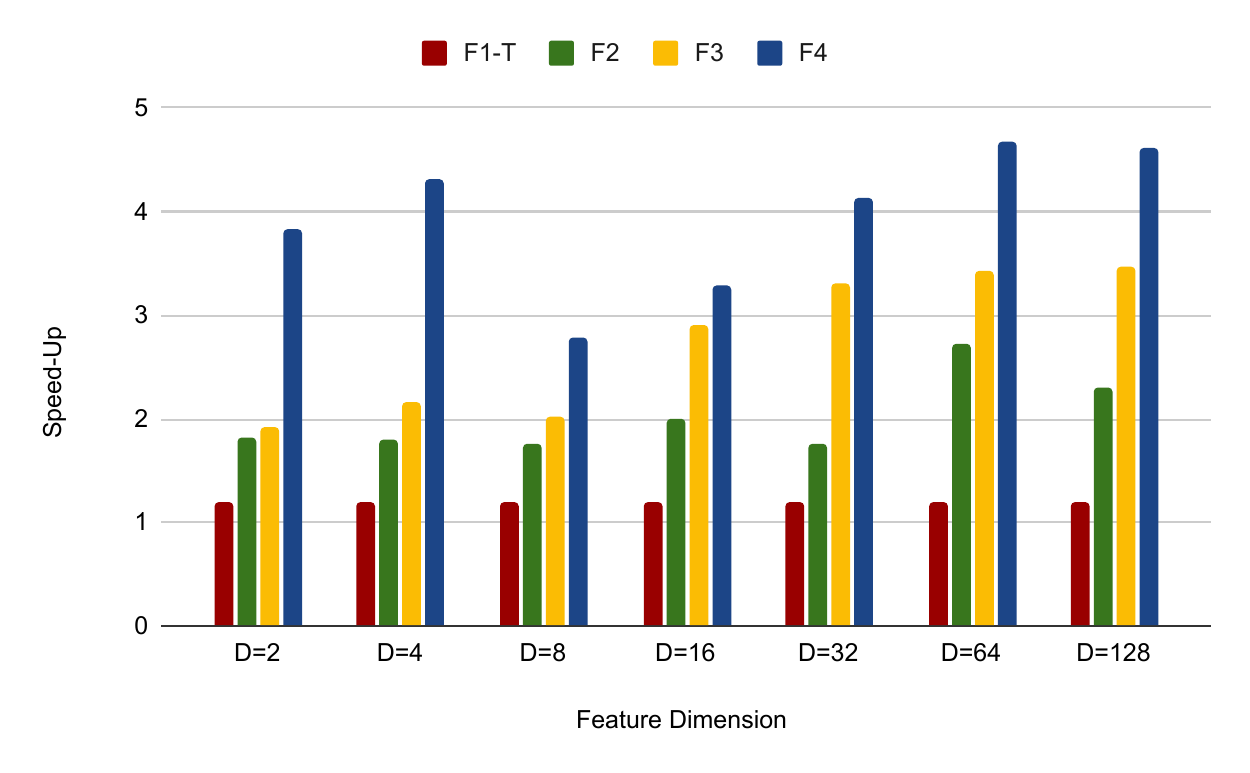}
  \caption{KNN: Latency speed-up of TAPA (F1-T), TAPA-CS (F2, F3, F4) compared with Vitis HLS baseline for K=10 and N=4M over varying feature size.}
  \label{fig:knn-dimension-latency}
\end{figure}

\begin{figure}
  \centering
  \includegraphics[height=5cm, keepaspectratio]{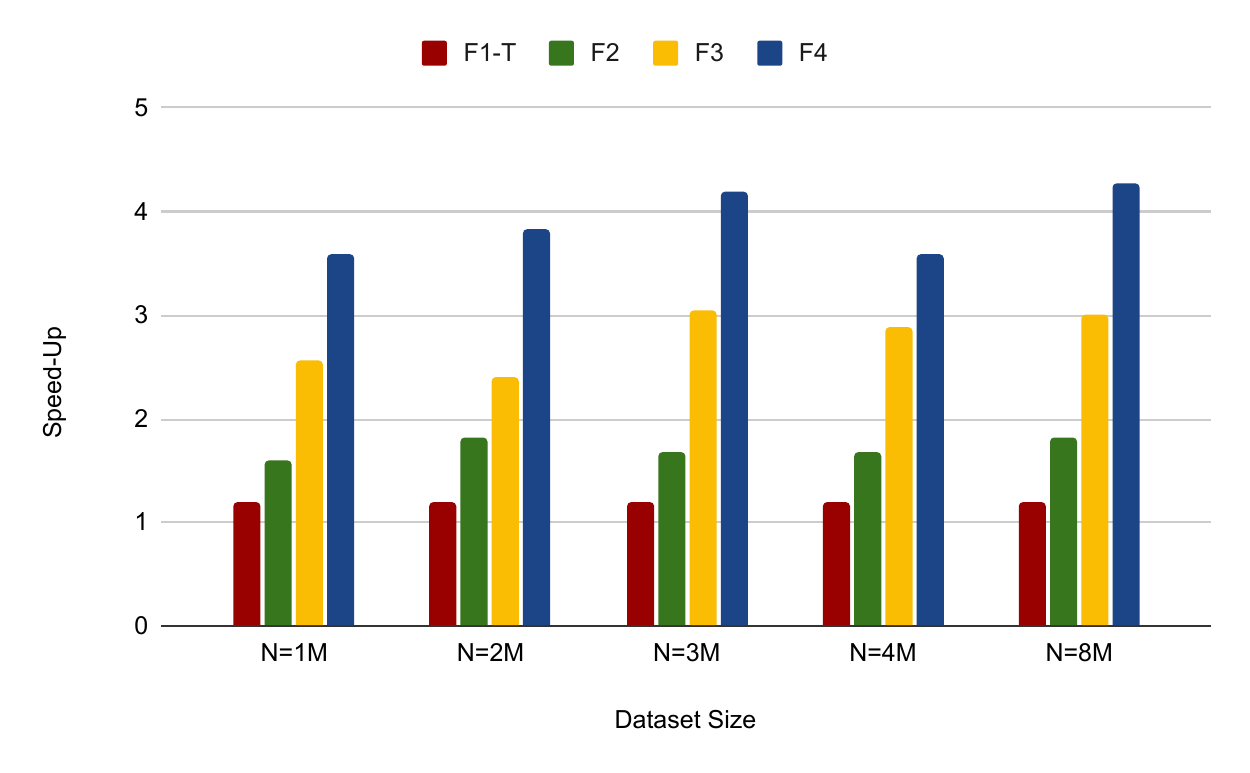}
  \caption{KNN: Latency speed-up of TAPA (F1-T), TAPA-CS (F2, F3, F4) compared with Vitis HLS baseline for K=10 and D=2 over varying dataset size.}
  \label{fig:knn-dataset-latency}
\end{figure}

Figure \ref{fig:knn-dimension-latency} presents the speed-up of the design generated by TAPA for a single FPGA and TAPA-CS over 2-4 FPGAs compared with the Vitis HLS baseline over varying feature dimensions. The 2, 3, and 4 FPGA designs produced by TAPA-CS are on average 2$\times$, 2.7$\times$, and 3.9$\times$ faster than the baseline Vitis version.
Next, we vary the dataset sizes and evaluate the performance of TAPA-CS over 2-4 FPGAs in Figure \ref{fig:knn-dataset-latency}. Compared with the Vitis-generated single-FPGA baseline, the 2, 3, and 4 FPGA designs are on average 1.7$\times$, 2.8$\times$, and 3.9$\times$ faster. Compared with the TAPA-generated single FPGA baseline, the designs are 1.4$\times$, 2.3$\times$, and 3.2$\times$ faster.

\begin{table}[h!]
\centering
\adjustbox{width=\columnwidth}{
\begin{tabular}{|c|c|}
\hline
\textbf{Parameters} & \textbf{Values} \\
\hline
$N$: Number of data points in the dataset & 1M, 2M, 3M, 4M, 8M \\
\hline
$D$: Dimension of the feature vectors & 2, 4, 8, 16, 32, 64, 128 \\
\hline
$K$ & 10 \\
\hline
\end{tabular}}
\caption{KNN parameters}
\label{knn-params}
\end{table}

\begin{figure}
  \centering
  \includegraphics[height=5cm, keepaspectratio]{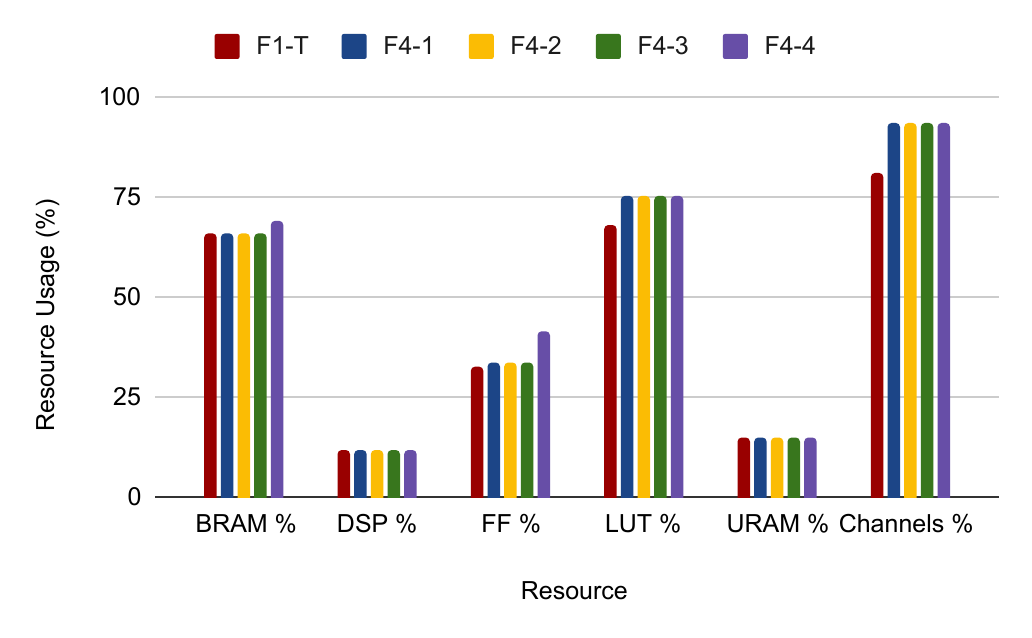}
  \caption{KNN: Resource utilization of single FPGA baseline (F1-T) and 4-FPGA design (F4). Here, F4-1 to F4-4 denote the 4 FPGAs used in design F4.}
  \label{fig:knn-resource}
\end{figure}

Figure \ref{fig:knn-resource} displays the resource utilization of the single FPGA baseline and each of the 4 FPGAs used in design F4. The resource utilization profiles of designs F2 and F3 are similar to that of F4 per FPGA. Therefore, we do not report them. 

The single FPGA baseline achieves a design frequency of 165MHz using Vitis, 198MHz using TAPA, and each of the 2, 3, and 4 FPGA designs generated by TAPA-CS achieve a design frequency of 220MHz displaying a 33\% increase compared with Vitis and 11.11\% increase compared with TAPA.
\subsection{CNN}
The CNN accelerator we chose is a systolic-array based implementation of the third layer of the VGG model \cite{simonyan2015deep} generated by AutoSA \cite{10.1145/3431920.3439292}. This accelerator consists of a grid of PEs performing identical computations. The grid size can be adjusted to meet throughput/resource constraints. 

The single FPGA baseline can successfully route a design with a grid size of 13x4 using Vitis and 13$\times$8 using TAPA. Larger designs like 13$\times$12, 13$\times$16, and 13$\times$20 either cause congestion or require more resources than what is available on a single device. We display the resource utilization profiles of these configurations in Table \ref{tab:cnn:resources}. Table \ref{tab:cnn-transfers} presents the inter-FPGA transfer volumes for the different grid configurations. In case of the CNN application, the compute intensity depends on the input size (54.5M floating-point operations), and the inter-FPGA transfer volumes increase with the increase in grid size.

\begin{figure}
  \centering
  \includegraphics[height=5cm, keepaspectratio]{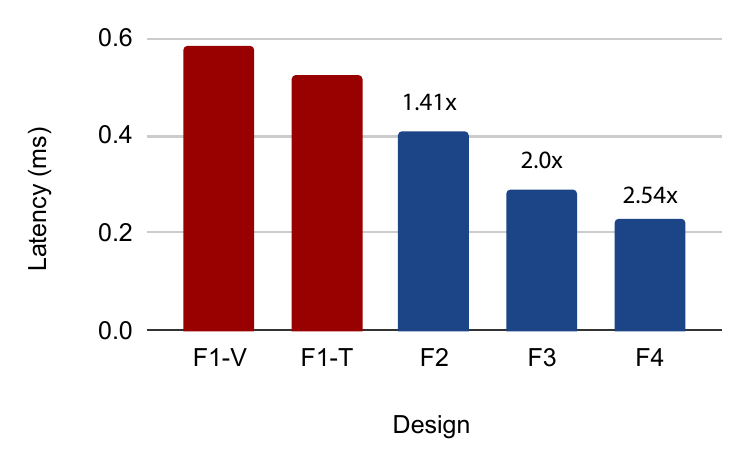}
  \caption{CNN: Latency comparison between TAPA-CS (F2, F3, F4), TAPA (F1-T), and Vitis-HLS (F1-V).}
  \label{fig:cnn-latency}
\end{figure}

We evaluate the single FPGA baselines (13$\times$4 through Vitis and 13$\times$8 through TAPA) against a 2 FPGA design with the 13$\times$12 grid, a 3 FPGA design with the 13$\times$16 grid, and a 4 FPGA design with the 13$\times$20 grid in Figure \ref{fig:cnn-latency}. Compared with the single FPGA Vitis baseline (13$\times$4), the 2-FPGA design (13$\times$12) is 1.41$\times$ faster, the 3-FPGA design (13$\times$16) is 2.0$\times$ faster, and the 4-FPGA design (13$\times$20) is 2.54$\times$ faster. There are two main reasons limiting the speed-up of CNNs when we scale to multiple FPGAs. First, as the grid size increases, the inter-FPGA data transfer sizes also increase (Table \ref{tab:cnn-transfers}). Second, due to the grid-like structure of the systolic-array, there are more PEs writing to AlveoLink requesting inter-FPGA data transfer than in the case of the other applications shown in Figure \ref{topo}. Due to the contention, there is a higher occurance of idle PEs on the second FPGA. 

\begin{table}[h]
    \centering
    \begin{tabular}{|c|c|}
    \hline
        Grid Size & Volume (MB) \\
        \hline
        13$\times$4 & 2.14 \\
        13$\times$8 & 4.28 \\
        13$\times$12 & 6.42 \\
        13$\times$16 & 8.57 \\
        13$\times$20 & 10.71 \\
        \hline
    \end{tabular}
    \caption{Inter-FPGA data transfer volumes over varying grid sizes for a fixed input size.}
    \label{tab:cnn-transfers}
\end{table}

The single FPGA baselines achieve a design frequency of 300MHz for 13$\times$4 using Vitis, and 13$\times$8 using TAPA, but they fail placement and routing for the larger grid sizes. TAPA-CS can successfully route grid sizes 13$\times$12, 13$\times$16, and 13$\times$20 on 2, 3, and 4 FPGAs achieving a design frequency of 300MHz on each FPGA.

\begin{table}
    \begin{tabular}{|c|c|c|c|c|c|}
\hline
Grid& LUT \% & FF\% &  BRAM\% & DSP\% & URAM\% \\
\hline
13$\times$4 & 20.4 & 12.1 & 14.2 & 25.2 & 0 \\
\hline
13$\times$8 & 38.3  & 23.5 & 23.7 & 49 & 0\\
\hline
13$\times$12 & 56.1 & 34.3 & 32.7 &  80.1 & 0 \\
\hline
13$\times$16 & 74 & 45.7 & 42.3 & 97.6 & 0 \\
\hline
13$\times$20 & 91.9 & 57 & 52.1 & 123.7 & 0 \\
\hline
\end{tabular}
    \caption{CNN: Resource utilization of different grid sizes. }
    \label{tab:cnn:resources}
\end{table}

\subsection{Overheads Added by TAPA-CS}\label{overheads_TAPA-CS}
TAPA-CS adds the following overheads to the overall HLS compilation stage and the resource utilization per FPGA:
\begin{enumerate}
    \item Floorplanning overheads added to the HLS compilation stage: We describe the floorplanning overheads added by the inter-FPGA floorplanning (L1) and intra-FPGA floorplanning (L2) for the Stencil and CNN applications below. Note that Stencil is the smallest application (15 compute modules) and CNN is the largest application (493 compute modules) tested in TAPA-CS. Therefore, the floorplanning steps add an overhead between 1.9s - 37.8s using Gurobi.

    \begin{tabular}{ cc }   
\textbf{Stencil} & \textbf{CNN} \\  
\begin{tabular}{ |c|c|c| } 
 \hline
      Iters & L1(s) & L2(s) \\
      \hline
      64 & 1.2&	0.7 \\
      \hline
      128 & 1.2 & 0.8 \\
      \hline
      256 & 1.2 &  0.8 \\
      \hline
\end{tabular} &  
\begin{tabular}{ |c|c|c| } 
\hline
Grid& L1(s)& L2(s) \\
\hline
13x4 & 0.3& 0.1\\
\hline
13x8 & 4.7& 2.7\\
\hline
13x12 & 14.7& 7.1\\
\hline
13x16 & 19.5& 9.3\\
\hline
13x20 & 24.6& 12.9\\
\hline
\end{tabular} \\
\end{tabular}

\item Resource overheads added by AlveoLink networking IPs: The networking IPs add the following negligible resource overhead per QSFP28 port per board:
    \begin{enumerate}
        \item LUT: 2.04\%
        \item FF: 2.94\%
        \item BRAM: 2.06\%
        \item DSP: 0\%
        \item URAM: 0\%
    \end{enumerate}
\end{enumerate}


\subsection{Scalability Beyond a Single Server Node}\label{scalability}
The experimental setup used to evaluate TAPA-CS consists of nodes with 4-FPGA rings. We have two such nodes containing a total of 8 FPGAs. To scale a design beyond a single node, we utilize host-side MPI and a 10Gbps ethernet link to transfer intermediate data. Note that this link is almost 10$\times$ slower than the connections between FPGAs on the same node. Table \ref{tab:hierarchy-data-transfers} describes the hierarchy of bandwidths involved in multi-FPGA design. We evaluate such an 8 FPGA setup for the following two applications:

\begin{enumerate}
    \item Stencil: We use the design with 512 iterations and scale the number of PEs to 120. As discussed in Table \ref{stencil-comp-inte}, each FPGA transfers a total data volume of 1153.73MB to the next FPGA. The overall runtime of the 8-FPGA setup is 11.65s, which is 1.45$\times$ slower than the single FPGA Vitis baseline. In this case, the main bottleneck is the high data movement, and sequential nature of the stencil application which leaves most of the FPGAs idle. The intermediate data from the first set of the 4-FPGA ring first needs to be transferred from the device memory to the host memory. Next, the data must be moved from one node to the other using the 10Gbps link. Lastly, the data has to be moved from the host memory of the second node to the device memory of the second 4-FPGA ring.  
    \item PageRank: We scale the PageRank application from 4 PEs (single FPGA) to 32 PEs for the 8-FPGA design and test it on the cit-Patents dataset described in Table \ref{tab:datasets-pagerank}. Note that unlike in the case of the Stencil application where each FPGA executes sequentially, in the case of PageRank (Figure \ref{topo}), once the vertex router module completes execution on the first FPGA, all other FPGAs can be launched in parallel. This increases the scalability of the design. The 8-FPGA design achieves an end-to-end latency of 3.44s, which is 1.4$\times$ faster than the single FPGA Vitis baseline. Note that despite the speed-up obtained in the case of PageRank, the 8-FPGA design is still slower than the 2-FPGA design on a single node. Therefore, the inter-node network adds significant latency, reducing the benefits of scaling. 
\end{enumerate}

\begin{table}[h]
\centering
\begin{tabular}{|c|c|}
\hline
Transfer & Bandwidth \\
\hline
On-chip (SRAM) & 35TBps \\
Off-Chip (HBM) & 460GBps \\
Inter-FPGA & 100Gbps \\
Inter-Node & 10Gbps \\
\hline
\end{tabular}
\caption{Hierarchy of data transfer bandwidths involved in multi-FPGA design.}
\label{tab:hierarchy-data-transfers}
\end{table}

\section{Related Works} \label{prior_work}

\subsection{Prior Work on Inter-FPGA Communication}
Prior attempts at leveraging the networking capabilities exposed by modern FPGAs can be classified into two categories. The first category of work uses host-orchestrated data transfers \cite{4100995, 9114837, 9651265}. In these works the host coordinates the inter-accelerator communication by exposing programmer-friendly MPI-like primitives. Using host-orchestration avoids re-programming the FPGA bitstream whenever a different communication pattern is to be followed. However, considering that several dataflow FPGA workloads suit streaming, where data is produced and consumed every cycle, host-orchestration adds significant overheads. The second category of works uses device-side initiation of inter-FPGA communication \cite{8541106, 10.1145/3295500.3356201,inproceedings}. Such papers benefit from the streaming nature of designs but suffer from frequent regeneration of the bitstream. 

We compare prior work and AlveoLink (described in Section \ref{networking-interface} in Table \ref{tab1}. Compared with EasyNet \cite{inproceedings} which also achieves a similar data transfer throughput of 90GBps, AlveoLink requires about half of the on-board resources. This allows larger designs to be mapped to the FPGA and utilize networking capabilities than in case of EasyNet. 
\begin{table}[htbp]

\begin{center}
\adjustbox{width=\columnwidth}{
\begin{tabular}{|c|c|c|c|}
\hline
Project & Orchestration& Resource Overhead (\%) & Performance (GBps) \\
\hline
TMD-MPI\cite{4100995}& Host&  26 & 10 \\
Galapagos\cite{8541106}& Device& 11.5 & 10 \\
SMI\cite{10.1145/3295500.3356201}& Device& 2 &40 \\
EasyNet\cite{inproceedings}& Device& 10 &90 \\
ZRLMPI\cite{9114837}& Host& - &10 \\
ACCL\cite{9651265}& Host& 16 &80\\
AlveoLink\cite{alveolink}& Device& 5&90\\
\hline
\end{tabular}}
\caption{Comparison of prior work addressing Challenge 1 in terms of data transfer throughput and FPGA resource area overhead. Here, "-" implies that the project does not discuss the area overhead.}
\label{tab1}
\end{center}
\end{table}

\subsection{Prior Work on Partitioning, Mapping, \& High Design Frequency}
There are some initial research efforts which address partitioning and mapping across multiple FPGAs which provide a good starting point for TAPA-CS, but they suffer from several shortcomings. Prior work such as Elastic-DF \cite{alonso} propose an ILP-based partitioner similar to ours, but suffer from poor frequency (190-240MHz) as they do not couple floorplanning and interconnect pipelining with HLS compilation. Also, Elastic-DF is integrated with the DNN inference compiler FINN \cite{Umuroglu_2017, blott2018finnr}, leading to poor generalizability to different workloads. 

A different approach to design partitioning leverages latency-insensitivity \cite{945302, 810667} to break the design at the latency-insensitive endpoints \cite{fleming, 10.1145/3373376.3378491, 9499729}. Such works view the design as consisting of multiple modules connected by FIFOs, allowing the authors freedom in implementing the inter-module communication. \cite{fleming} leverage this technique, but expect the user to provide a static module-to-FPGA mapping, use Verilog/VHDL for specifying the modules, and perform simulation-based experiments. In case of TAPA-CS, we partition the design at latency-insensitive endpoints and automatically find the optimal module-to-FPGA mapping, expose an easy-to-use C++ interface to the user, and perform experiments on real FPGAs achieving high frequency. Virtualization-based works such as \cite{10.1145/3373376.3378491, 9499729, 10.1145/3445814.3446699} assign modules to pre-placed and pre-routed "soft blocks" which does not scale well to real-world large-scale designs where each function is compiled into an RTL controlled by a finite state machine (FSM) and has varying resource requirements. 

SMAPPIC \cite{10.1145/3575693.3575753} introduces a multi-node emulation system where each node can be a single die of an FPGA or the whole FPGA. It uses the computational cores shipped with BYOC \cite{10.1145/3373376.3378479} assign cores to the nodes. However, SMAPPIC uses Gen3x16 PCIe-based connections between FPGAs which provides a slow round-trip latency of 1250ns, and does not provide the tool with a hardware layout of the FPGAs. This leads the designs to have a low final frequency of 100MHz. In contrast, TAPA-CS has an inter-FPGA round trip latency of 1 $\mu$s (12.5x faster than SMAPPIC), provides the partitioning tool with a global view of the chip layout allowing us to achieve a high frequency of between 266-300MHz. We also do not fix the granularity of resource allocation for a computational core/module to a node, allowing greater flexibility in the target applications. For example, in TAPA-CS a single die can contain any number of modules, and modules spanning across multiple dies are pipelined sufficiently to maintain the final frequency.

\section{Challenges in Multi-FPGA Design}
Modern FPGAs are well-equipped to support scalable accelerator designs spanning across multiple devices and nodes as displayed in Section \ref{eval}. However, compared with the efforts in scalable design across CPUs and GPUs, the FPGA community still has several open challenges to address. We outline some of them below:

\begin{enumerate}
    \item Workload perspective: Most FPGA designers rely on domain-specific knowledge to produce accelerators optimized to the resources available on a single device. That is, conventionally, designers create kernels using at most 32 HBM channels even if their design is memory-bounded, and decide tiling and unroll factors based on available on-chip resources. However, with network-connected multi-FPGA environments, FPGA designers need to think beyond a single FPGA and scale their designs accordingly. CPU-based designs can be scaled easily through multi-threading and GPU-based designs offer data-parallel and model-parallel modes through the PyTorch front-end which make it easier for users to design large applications without considering the physical constraints of each device. TAPA-CS provides a partitioning framework for large scaled-up designs. However, there is a lack of frameworks which automatically enable scaling-up a design from a single FPGA to multiple FPGAs. We are currently working on map-reduce style programming frameworks for FPGAs which will allow automated scaling based on the memory-/compute- intensity of the application, combined with the partitioning introduced in this paper. 
    \item Inter-FPGA communication perspective: As discussed in Section \ref{prior_work}, compared with other open-sourced communication protocols, AlveoLink offers the best theoretical latency (1$\mu$s) at the lowest resource budget. However, in practice, the latency varies greatly with different packet sizes (minimum transfer unit) and total volume of data. For example, a data transfer of 64 MB with packet size of 64 bytes takes a total of 6.53 milliseconds, while the same volume with a packet size of 128 bytes takes a total of 3.96 milliseconds. This overhead can be significant for applications with strict latency constraints. Ideally, FPGA-based communication should add an overhead in the order of <100ns to be widely adopted for different applications. We are working with FPGA vendors to collaborate on such a solution.
\end{enumerate}

\section{Conclusion}
This paper presents TAPA-CS - a task-parallel dataflow programming framework which automatically partitions, generates, and compiles a large design across a cluster of FPGAs achieving high throughput and frequency. TAPA-CS uses an ILP-based partitioning framework which takes into account resource utilization profiles of compute modules before mapping them to different FPGAs. It also couples a coarse-grained floorplanning step with pipelining at the inter- and intra-FPGA levels to ensure high accuracy. We test the design across multiple benchmarks of varying compute and memory requirements to validate the scalability and generalizability of the tool. Across all the tested designs, TAPA-CS achieves an average throughput improvement of 2.1$\times$, 3.2$\times$, and 4.4$\times$ using 2, 3, and 4 FPGAs respectively, and frequency improvement between 11-116\% compared with single FPGA designs generated through Vitis HLS. We plan to open-source TAPA-CS once the paper is accepted, which will allow us or the community to add support for Intel FPGAs in the near future. 

\bibliographystyle{plain}
\bibliography{references}

\end{document}